\documentclass{appolb}
\usepackage{graphicx}
\usepackage{amsmath}
\usepackage{amsfonts}
\usepackage{amssymb}
\usepackage[table]{xcolor}
\usepackage{array}
\usepackage[skip=3pt]{caption} 
\usepackage[skip=-0.3em]{subcaption}
\usepackage[pagewise]{lineno}							  %

\begin{document}
\title{The study of strongly intensive observables\\ for $\pi^{\pm,0}$ in $pp$ collisions at LHC energy in the framework of \\ PYTHIA model%
}
\author{Tumpa Biswas, Dibakar Dhar, Azharuddin Ahmed, \\Prabir Kr. Haldar\footnote{prabirkrhaldar@gmail.com (corresponding author)}
\address{Department of Physics, Cooch Behar Panchanan Barma University, \\ Cooch Behar - 736101, India}
\\[3mm]
{Abdel Nasser Tawfik 
\address{Future University in Egypt (FUE), Fifth Settlement, End of 90th Street, 11835 New Cairo, Egypt}
}
}
\maketitle

\begin{abstract}
The fractal and phase transitional properties of each type of pions (i.e. $\pi^{\pm,0}$) through one-dimensional $\eta-$space, at an energy of $\sqrt{s}=13~$TeV, \textcolor{black}{have} been studied with the help of the Scaled Factorial Moment (SFM) framework. To generate simulated data sets for $pp$ collisions under the minimum bias (MB) condition at $\sqrt{s}=13~$TeV, we have employed the Monte Carlo-based event simulator PYTHIA. 
Various parameters such as the Levy index $(\mu)$, degree of multifractality $(r)$, anomalous fractal dimension $(d_q)$, multifractal specific heat $(c)$ and critical exponent $(\nu)$ have been calculated. \textcolor{black}{To study the Bose Einstein(BE) effect due to identical particles (here pions) we have also derived these parameters for mixed pion pairs (i.e. $\{\pi^{+},\pi^{-}\}$, $\{\pi^{+},\pi^{0}\}$ and $\{\pi^{-},\pi^{0}\}$) and we find that the effects of identical particles weakened for the mixture with respect to the individual distributions.} The quest for the quark-hadron phase transition has also been conducted within the framework of the Ginzburg-Landau (GL) theory of second-order phase transition. Analysis revealed that for PYTHIA-generated MB events, there is a clear indication of the quark-hadron phase transition according to the GL theory. Furthermore, the values of the multifractal specific heat ($c$) for each $\pi^{+}, \pi^{-}, \pi^{0}$ and \textcolor{black}{the mixture pair data sets of pions} generated by PYTHIA model at MB condition, indicate a transition from multifractality to monofractality in $pp$ collisions at $\sqrt{s}=13~$TeV.
\end{abstract}
  
\section{Introduction}
Scientists are investigating the interactions of heavy ions at high temperatures and densities to gain insights into the evolution of multiple particle production. Their main focus lies in investigating the creation of quark-gluon plasma (QGP), which is a distinct phase of nuclear matter. Studying multiplicity fluctuations, specifically density fluctuations in charged particles, may offer a possible indication of the creation of a Quark-Gluon Plasma (QGP) \cite{1a,1b,1c,1d,1e,1f}. Pions are subatomic particles in particle physics that have the lowest energy. They are formed of one quark and one antiquark, and they are known to be unstable. The charged pions undergo decay into muons and muon neutrinos with an average lifetime of $2.60\times 10^{-8}$ seconds, while the neutral pion decays into gamma rays within a much shorter time frame of $8.5\times 10^{-17}$ seconds. Pions, along with other vector mesons such as $\rho$ and $\omega$, are hypothesized to account for the residual strong force that exists between nucleons. They are commonly formed during high-energy collisions between subatomic particles and play a crucial role in the annihilation of matter and antimatter. Pions, which have a spin of zero, consist of quarks from the first generation. In the quark model, the fusion of an up quark and an anti-down quark produces the $\pi^+$ particle, while a down quark combined with an anti-up quark generates the $\pi^-$ particle \cite{1f1,1f2}. These two particles are considered antiparticles of each other. The neutral pion $\pi^0$ is created through the combination of an up quark with an anti-up quark or a down quark with an anti-down quark. It is crucial to acknowledge that these two combinations exist in superpositions as a result of their equal quantum numbers. Out of the three pion mesons, the $\pi^0$ meson, which is both its own antiparticle and in a superposition state, has the lowest energy level. 

The study determined the anomalous fractal dimension $(d_q)$, which is a parameter utilized to examine the fractality of data. 
The findings reveal that an elevation in $d_q$ signifies the presence of multifractality, whereas a consistent value indicates monofractality. Phase transitions play a critical role in the study of particles, and theories such as Ginzburg and Landau's\cite{1g} phenomenological theory offer valuable insights into systems of condensed matter. The application of non-linear optics has been \textcolor{black}{utilized} to study the phase transition from hadrons to quarks in ultra-relativistic heavy-ion interactions. This investigation incorporates Hwa's relationship between fractality and intermittency theories. Scale factorial moments can detect non-statistical fluctuations, which can be important indications of the quark-hadron phase shift. Therefore, investigating fluctuations is crucial in understanding this transition. \textcolor{black}{Higher order correlations have been observed for cosmic ray, $e^+e^-$, nucleus-nucleus, hadron-hadron, and lepton-hadron interactions as particle-density fluctuations. To study these fluctuations in detail, normalized factorial moments have been analyzed and provided evidence for a correlation effect self-similar over a large range of the resolution, known as intermittency. Intermittency is found to be all-present in hadron production and is evidence for genuine correlations to high orders, but it seems dominated by Bose-Einstein(BE) correlations. At lower energies, it has been noted that the normalized factorial moments (or correlation integrals) in hadron-hadron collisions exhibit an almost linear relationship with the rapidity interval and the findings support the significant impact of identical particle correlations on the factorial moments and their scaling behavior \cite{1m,1n,1o}.} 
Phase transition analysis entails the examination of the ratio between upper and second-order fractal dimensions, which is represented as $\beta_q=(q-1)^\nu$. The critical exponent $\nu$ offers insights into the phase transition between hadrons and quarks. If the value of $\nu$ roughly matches the critical value, a phase change from quark to hadron may occur. This study combines non-statistical fluctuations with the SFM framework to analyze one-dimensional $\eta$-space collisions at an energy of $\sqrt{s} =$ 13 TeV. The study investigates minimum bias events by utilizing the Monte Carlo event generator PYTHIA. The main focus is on analyzing the behaviour of fractal moment fluctuation and intermittency index.

\section{Goal of the study}
\label{sec:goal}
Scientists from around the world have carried out multiple tests at CERN, RHIC, and LHC, confirming the formation of a state of Quantum Chromodynamics where colours are no longer restricted \cite{1h,1i,1j,1k,1l}. At high energies, collisions between protons (pp collisions) can yield valuable insights into the interactions of heavy ions. There is a particular focus on investigating pp collisions and occurrences with a higher number of particles involved. Understanding $AA$ interactions at the RHIC and LHC energies requires a clear depiction of $pp$ interactions at relativistic energies. Transverse momentum spectra estimate charged particle numbers in $pp$ collisions, providing input for theoretical models. Studying $pp$ collisions in depth helps understand system properties and phase changes. Comparing charged particle multiplicities with peripheral high energy aspects of $AA$ interactions is interesting.

\textcolor{black}{As early mentioned at lower energies the intermittency effects are explained by the BE correlations between like-signed pions\cite{1m,1n,1o,1p}. In this analysis, we want to study the effects of BE correlation on intermittency at ultra relativistic energies by comparing the results of $\pi^+, \pi^- , \pi^0$ with their pairs (i.e. $\{\pi^{+},\pi^{-}\}$, $\{\pi^{+},\pi^{0}\}$ and $\{\pi^{-},\pi^{0}\}$). Expected output will be the effect of identical pions should disappear or be weakened for the mixture concerning the individual distributions.}

Previously, a study has already been done on the multiplicities of pions generated in $pp$ interactions at 13 TeV \cite{Shreya,Shreya1}. 
This examination covered the pseudo-rapidity $(\eta)$, azimuthal angle $(\phi)$, and $\eta-\phi$ phase spaces, employing precise topological parameters. Some of these parameters include the degree of multifractality $(r)$, the anomalous fractal dimension $(d_q)$, the Levy index $(\mu)$, the critical exponent $(\nu)$, and the multifractal specific heat $(c)$. We have already discussed that $\pi^+$, $\pi^-$ and $\pi^0$ exhibit distinct quark structures and lifetimes, our interest lies in gaining a deeper understanding of each type of pions. So in this motive, we explore separately the nature of fractality and phase transition for each type of pions, i.e. $\pi^+$, $\pi^-$ and $\pi^0$ within the framework of Scaled Factorial Moment (SFM) -- from PYTHIA simulated $pp$ collisions at a particular impact parameter and MB condition at an energy of $\sqrt{s}=$13 TeV along the pseudorapidity space. In this investigation, we have utilized simulated datasets of $\pi^+$, $\pi^-$ \& $\pi^0$ and \textcolor{black}{pion pair data sets of $\{\pi^{+},\pi^{-}\}$, $\{\pi^{+},\pi^{0}\}$ and $\{\pi^{-},\pi^{0}\}$ }derived from $pp$ collisions at $\sqrt{s}=13$ TeV, generated by PYTHIA model under MB condition. The main goal is to record intermittent variations in the density of particle distribution.

\section{Data Description} 
\label{sec:data}
\subsection{Experimental Details}
\label{sec:expr}
 In this investigation, we have utilized datasets of $\pi^{\pm,0}$ particles \textcolor{black}{along with their mixture pairs (i.e. $\{\pi^{+},\pi^{-}\}$, $\{\pi^{+},\pi^{0}\}$ and $\{\pi^{-},\pi^{0}\}$)} from the PYTHIA model to analyze the fluctuations in particle density distribution within $pp$ collisions at an energy of $\sqrt{s}=13$ TeV. The analysis encompassed simulated MB datasets as in real experiments, determining impact parameters for $pp$ interactions at LHC energies is challenging. The transverse momentum ($p_T$) distribution of pions derived from the  PYTHIA-simulated datasets at MB condition have been compared in the present work. This analysis was carried out in Fig. 1 in conjunction with the ALICE experimental MB data sets \cite{ALICE2020}. It suggests that the model \textcolor{black}{exhibits} a qualitatively similar trend to those observed in the experiments.

\begin{figure}[h]
\begin{center}
\captionsetup{justification=centering}
\includegraphics[width=0.77\textwidth]{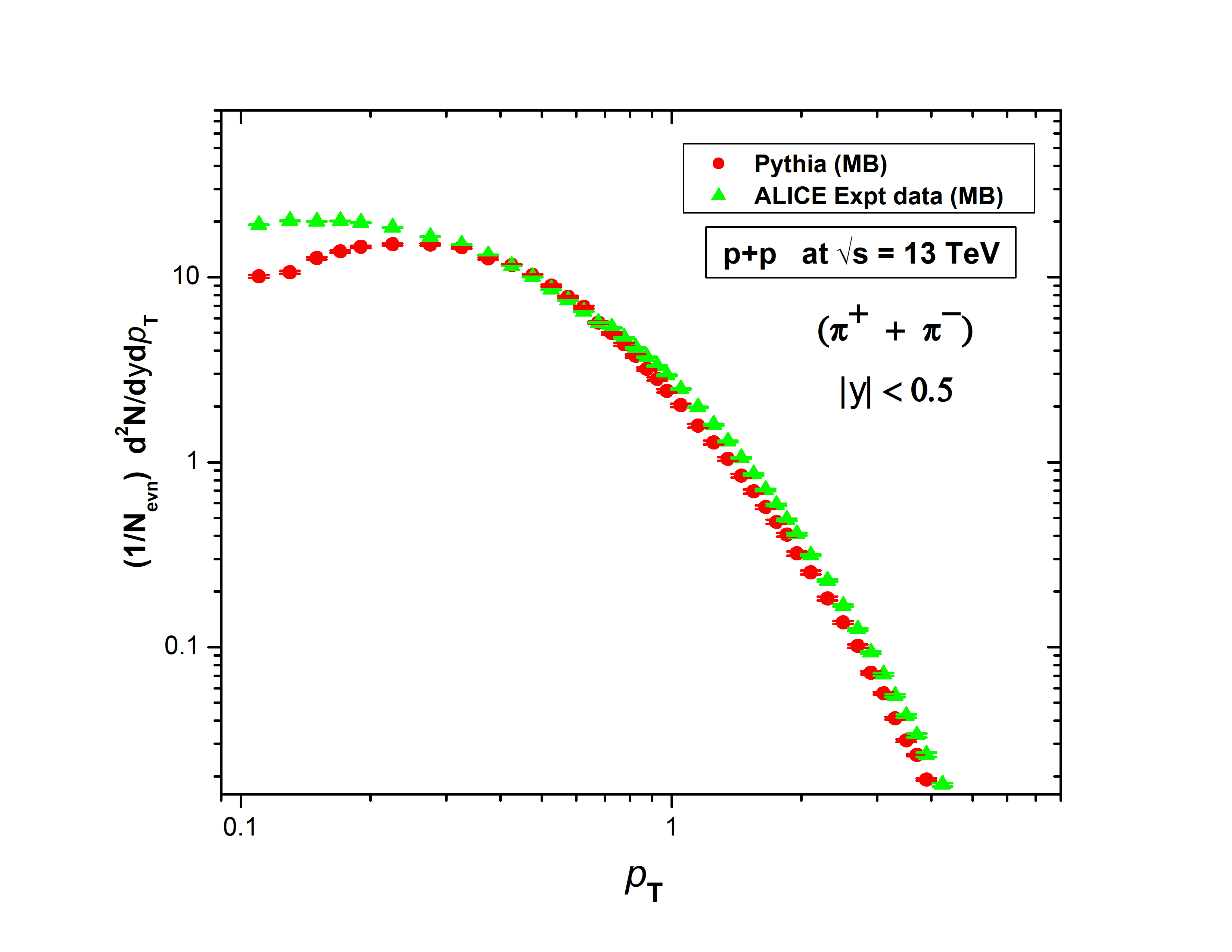} 
\caption{Relative analysis of the transverse momentum ($p_T$) distribution of pions produced from the PYTHIA-generated datasets at MB condition. This analysis was carried out in conjunction with the ALICE experimental MB data sets at $\sqrt{s} = 13$ TeV. \cite{ALICE2020}.}
\end{center}
\end{figure}

\subsection{PYTHIA Simulation}
 \label{sec:pythia}
PYTHIA is a widely utilized event generator for the analysis of collisions involving protons and leptons, specifically $(pp)$ and proton-lepton interactions. The latest progress in  PYTHIA allows for the investigation of high-energy collisions involving heavy atomic nuclei, specifically $pA$ \& $AA$ interactions. In this study, we utilize the PYTHIA event generator to simulate pp collisions at 13 TeV.  We employ  PYTHIA \cite{Sjostrand2008} version 8.3, which incorporates multi-parton interactions (MPI). MPI is crucial for elucidating the fundamental processes, Multiplicity distributions, and generation of charmonia. Typically, a high-energy event generator generates between four to ten interactions between partons, which are influenced by the overlapping area of the colliding particles \cite{Weber2019}. Initial State Radiation, abbreviated as ISR \& Final State Radiation, also termed as FSR \cite{Martinez2014,Lees2015} are used to accomplish the perturbative scattering processes. The PYTHIA v8.3 framework is structured into three primary components: hadron level, process level and parton level  \cite{Egede2022}. The hard-scattering process, which generates temporary resonances, is depicted at the process level. Usually, the hard process is perturbatively expressed, involving a restricted number of particles, often confined within high-energy intervals\cite{Sjostrand2015}. At the parton level, there are several shower models available, including both the preliminary and final state radiation. Currently, the analysis includes the consideration of multiparton interactions, as well as the handling of the remnants of \textcolor{black}{the} beam and the possibility of colour reconnection occurrences.\\

The PYTHIA simulation program uses the Lund string fragmentation model to achieve hadronization. The Lund area law describes the probability of hadron production. The residual beams and partons are coupled through potential energy strings. Strings rupture, generating additional quark-antiquark pairs. This process fragments the strings into shell hadrons, reducing particle production and multiplicity\cite{Andersson2001,Ortiz2013}. 
The coherence in final-state particles cannot be directly described because of the probabilistic nature of the consideration of the phenomenological models of the hadronization in PYTHIA. Bose-Einstein (BE) effects, where correlations arise between identical bosons in an event from symmetrization of the production amplitude, are also a classical example of these type of final state of coherence. Although these correlations are expected to have a negligible impact for most measurements in pp collisions but the BE effects have been observed in minimum bias pp collisions \cite{aa1,aa2}. The intermittency effects can be partly or totally due to the BE effect due to identical particles, here identical pions \cite{aa3}. In this analysis, we are working with $\pi^+,\, \pi^-$ and $\pi^0$ but in PYTHIA v8.3 the BE Correlations are not considered by default, so during data generation we have considered these effects.


\section{Method of Analysis}
\label{sec:method}
In accordance with Bialas and Peschanski, in a 1-dimensional phase space partitioned into $M$ bins, the factorial moment of order $q$, designated as $F_q$, can be represented as follows \cite{2,3}:
 \begin{equation}
 F_q = M^{q-1}\sum_{m=1}^M \frac{<n_m(n_m-1)\cdots(n_m-q+1)>}{n_m^q},
 \end{equation}
Here, $n_m$ is the total number of particles within the m-th bin, and $q$ denotes the order of the moment. The symbol $\langle ....\rangle $ denotes the event average. 
 
 As previously mentioned, $F_q$ exhibits a power-law relationship with respect to $M$ \cite{34a}, specifically $\langle F_{q}\rangle \propto M^{\alpha_q}$, or in logarithmic form:

\begin{equation}
\ln \langle F_q\rangle= \alpha_q \; \ln M + A
 \end{equation}
 
This phenomenon is commonly referred to as ``intermittency'' \cite{34b,34c}, with $\alpha_q$ denoting the intermittency strength, also known as the intermittency exponent. Here, $A$ representing a constant, and Eqn.~(2) can be used to evaluate $\alpha_q$ by best fit analysis. 

Additionally, the relationship between the anomalous fractal dimension $d_q$ and the intermittency exponent $\alpha_q$ has been established \cite{34c,34d}:
 
 \begin{equation}
 d_q=\frac{\alpha_q}{q-1},
 \end{equation}
Here, $d_q$ represents the Renyi co-dimension. Lastly, we define $D_q$, the universal fractal dimension, as follows: 

 \begin{equation}
 D_q= (1-d_q).
 \end{equation}
 
\subsection{Critical Exponent}  
\label{sec:crtclexp}

Ochs's \cite{ochs} simple scale-invariant cascade model predicts that higher-order scale factorial moments are related to second-order scaled factorial moments using a modified power law equation:

\begin{equation}
F_q \propto F_2^{\beta_q},
\end{equation}

The relation holds the potential to yield essential insights into the underlying dynamics of the system. Interestingly, it has been observed that the slopes of the power law coupling higher-order and second-order SFM remain constant regardless of the size of the phase space and phase dimension \cite{Hwa,Hwa1}. In essence, the values of $\beta_q$ encapsulate the system's scale-invariant behavior on a global scale. The $\beta_q$ values are generated from the ratio of the higher-order anomalous fractal dimension $d_q$ to the 2nd-order anomalous fractal dimension $d_2$, as described in the equation below:

 \begin{equation}
 \beta_q = \frac{d_q}{d_2}(q-1).
 \end{equation}
Furthermore, the parameter $\nu$, known as `critical exponent' can be determined through the relationship:
 \begin{equation}
 \beta_q= (q-1)^\nu.
 \end{equation}
 
According to the Ginzburg-Landau (GL) hypothesis, if the value of $\nu$ closely matches to $1.304$ in a certain dataset, it indicates the presence of a hadron-quark phase transition. If the value of $\nu$ differs greatly from $1.304$, it implies the absence of a hadron-quark phase transition.

 \subsection{Levy Index}
 \label{sec:levy}
The Levy stable laws, originally established by Brax \& Peschanski for the examination of intermittency in high energy heavy-ion interactions \cite{38a}, rely on the Levy stability index $\mu$ to gauge the degree of multifractality in the multiparticle creation process. This index $\mu$ exhibits a continuous spectrum within the stability range of $0\leq\mu\leq 2$.When $\mu$ value is equal to $2$, it denotes the existence of minimal fluctuations, while a zero value of $\mu$ suggests the highest level of fluctuations within the self-similar mechanism. If the value of $\mu$ belongs to $0$ and $1$, a thermal phase transition is suggested, while $\mu>1$ implies the potential existence of a non-thermal phase changes in the cascading machanism process. According to the Levy law, the ratio of anomalous dimensions depends on $\mu$ and can be expressed through the following relationship \cite{38a,38b}: 
 
 \begin{equation}
 \beta_q=\frac{(q^\mu-q)}{(2^\mu-2)}.
 \end{equation}
 
However, the $\mu$ is described as the degree of multifractality (whereas, $\mu=0$ for monofractals). Values of $\mu<1$ indicate ``calm'' singularities, while $\mu>1$ relate to ``wild'' singularities. 
 
\subsection{Multifractal Specific Heat}
\label{sec:specifcheat}
Bershadkii \cite{39} proposed the notion of constant heat approximation, which is often used to explain the specific heat of gases and solids at a constant temperature, to be relevant in multifractal data analysis of multipion production processes. In this specific context, when the condition $D_q$ surpasses $D_q'$ for $q$ less than $q'$ with respect to the multiplicity of charged particles, it implies the existence of multifractality. By relying the generalized fractal dimension, the determination of multifractal specific heat, denoted as $c$, can be achieved through the following expression\cite{39,40}:

 \begin{equation}
 D_q=(a-c)+\left[c \times \left(\frac{\ln q}{q-1}\right)\right],
 \end{equation}
In this context, where $c$ signifies the specific heat, a positive value for $c$ substantiates the occurrence of a phase transition from multifractality to monofractality in the case of heavy-ion interactions\cite{41,42,43,44}. 
 
\section{Result and Discussion}

\label{sec:results}
To examine phase transitions and fractality using SFM, we have considered the PYTHIA simulated $pp$ collisions for each type of pions, i.e., $\pi^{\pm,0}$ and \textcolor{black}{pion pair data sets of $\{\pi^{+},\pi^{-}\}$, $\{\pi^{+},\pi^{0}\}$ $\&$ $\{\pi^{-},\pi^{0}\}$ } at an energy of $\sqrt{s}=13~$TeV  in $\eta$ (pseudorapidity) -space.

\begin{figure*}[h!]
\begin{center}
\captionsetup{justification=centering}
\includegraphics[scale=1.3]{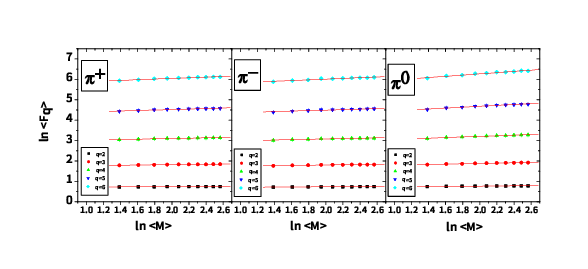} 
\caption{Plot of $\ln<F_q>$ vs. $\ln M$ in $\eta$-space for $\pi^+$, $\pi^-$ \& $\pi^0$ for PYTHIA generated MB events. \textcolor{black}{The lines joining the data points are the best fit lines.}}
\end{center}
\end{figure*}

\begin{figure*}[h!]
\begin{center}
\captionsetup{justification=centering}
\includegraphics[scale=1.2]{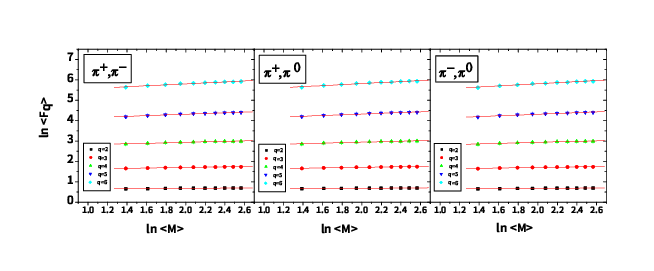} 
\caption{ \textcolor{black}{Plot of $\ln<F_q>$ vs. $\ln M$ in $\eta$-space for $\{\pi^{+},\pi^{-}\}$, $\{\pi^{+},\pi^{0}\}$ and $\{\pi^{-},\pi^{0}\}$ for PYTHIA generated MB events. The lines joining the data points are the best fit lines.}}
\end{center}
\end{figure*}



\begin{figure}\centering \captionsetup{justification=centering}
\subfloat[]{\label{a}\includegraphics[width=.45\linewidth]{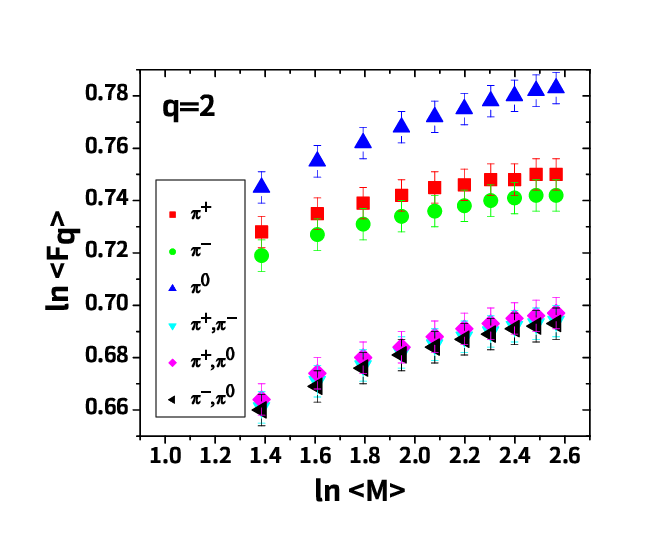}}\hfill
\subfloat[]{\label{b}\includegraphics[width=.45\linewidth]{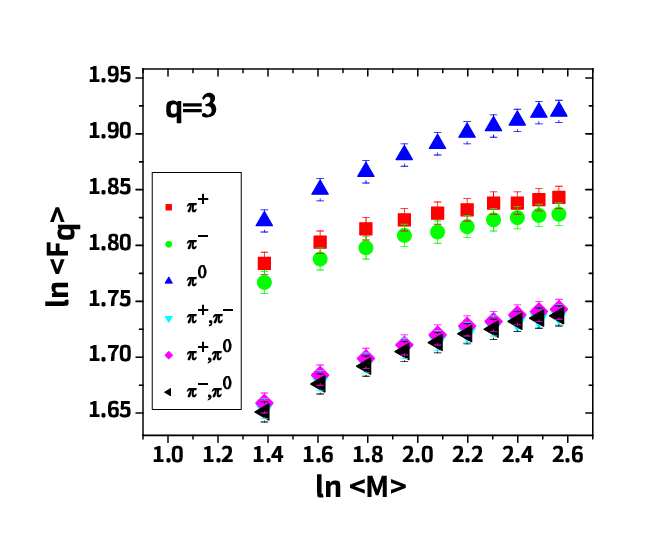}}\par 
\subfloat[]{\label{c}\includegraphics[width=.45\linewidth]{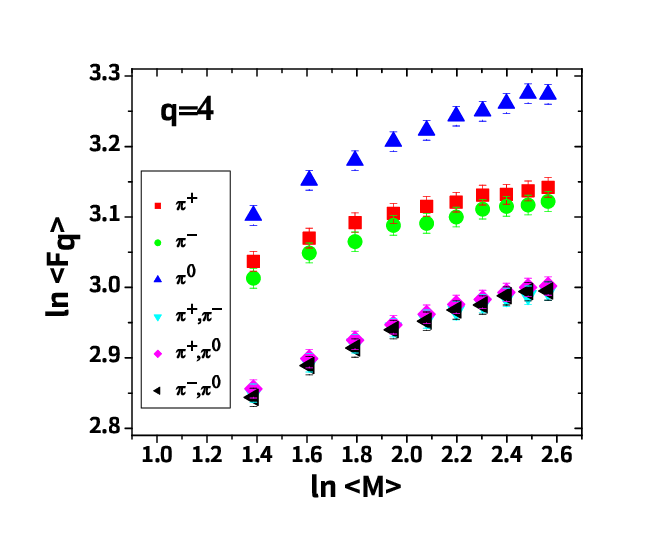}}
\caption{\textcolor{black}{Comparison plot of $\ln<F_q>$ vs. $\ln M$ in $\eta$-space for $\pi^+$, $\pi^-$, $\pi^0$ with $\{\pi^+,\pi^-\}$, $\{\pi^+,\pi^0\}$ $\&$ $\{\pi^-,\pi^0\}$}}
\label{fig}
\end{figure}
 
As suggested by Bialas and Peschanski \cite{2,3}, we have estimated the values of the factorial moment $F_q$ for various orders $q$, considering different numbers of bins $M=2,3,4,\cdots,20$ in 1-dimensional $\eta$-space by using the relation $\langle F_q\rangle \propto M^{\alpha_q}$. We have conducted the analysis of the variation of $\ln \langle F_q\rangle$ versus $\ln \langle M\rangle$ for different orders of $q$ in one-dimensional $\eta$-space. Fig.~2 and Fig.~3 has depicted the relation between $\ln \langle F_q\rangle$ and $\ln \langle M\rangle$ for PYTHIA-generated events. \textcolor{black}{Whereas Fig.~4 represents a comparison of $\pi^+, \pi^- \text{ and } \pi^0$ with their pairs (i.e. $\{\pi^{+},\pi^{-}\}$, $\{\pi^{+},\pi^{0}\}$ and $\{\pi^{-},\pi^{0}\}$) for the variation of $\ln \langle F_q\rangle$ versus $\ln \langle M\rangle$ for different orders of $q$. From Fig.~4 it is clearly visible that the variation of $\ln \langle F_q\rangle$ versus $\ln \langle M\rangle$ is quite large compared to the mixture pairs. Also, it is notable that for positively charged pions these variations are even larger than the negatively charged and neutral pions. Although from Fig.~2 and Fig.~3 it is clearly visible that the variation of $\ln \langle F_q\rangle$ is almost uniform with $\ln \langle M\rangle$ throughout the region for individual pions and also for the mixture pairs.}

\begin{figure}
     \centering
     \captionsetup{justification=centering}
     \begin{subfigure}[b]{0.49\textwidth}
         \centering
         \includegraphics[width=\textwidth]{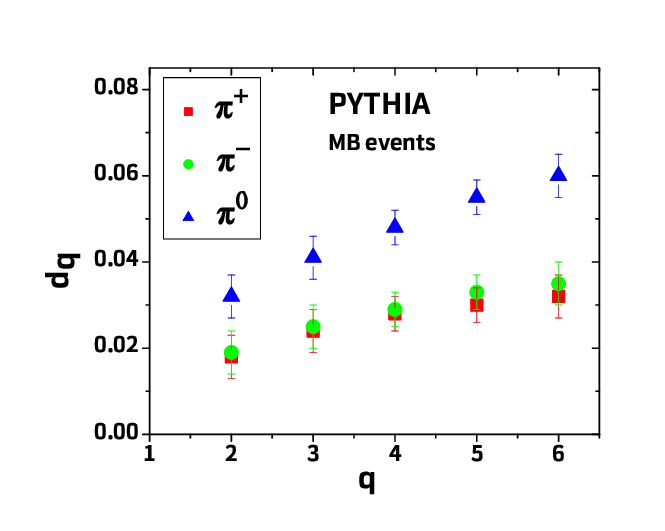}
         \caption{}
     \end{subfigure}
     \hfill
     \begin{subfigure}[b]{0.49\textwidth}
         \centering
         \includegraphics[width=\textwidth]{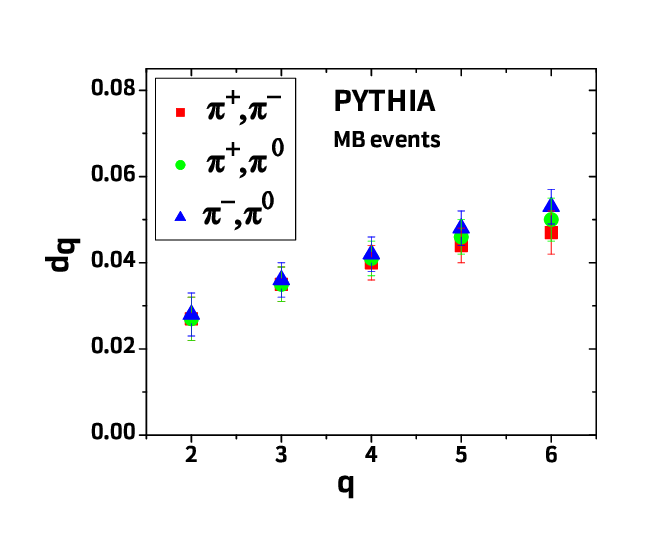}
         \caption{}
     \end{subfigure}
        \caption{\textcolor{black}{Plot of $d_q$ vs. $q$ in $\eta$-space: (a) for $\pi^+$, $\pi^-$ \& $\pi^0$ and \\ (b) $\{\pi^+,\pi^-\}$, $\{\pi^+,\pi^0\}$ $\&$ $\{\pi^-,\pi^0\}$.}}
\end{figure}






\begin{figure}
     \centering
     \captionsetup{justification=centering}
     \begin{subfigure}[b]{0.49\textwidth}
         \centering
         \includegraphics[width=\textwidth]{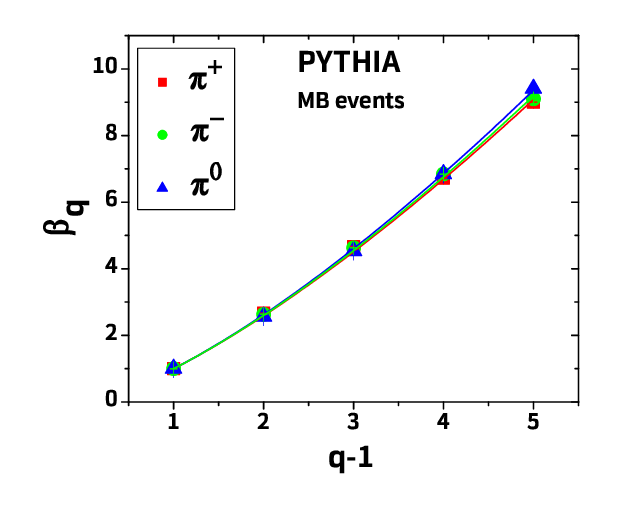}
         \caption{}
     \end{subfigure}
     \hfill
     \begin{subfigure}[b]{0.49\textwidth}
         \centering
         \includegraphics[width=\textwidth]{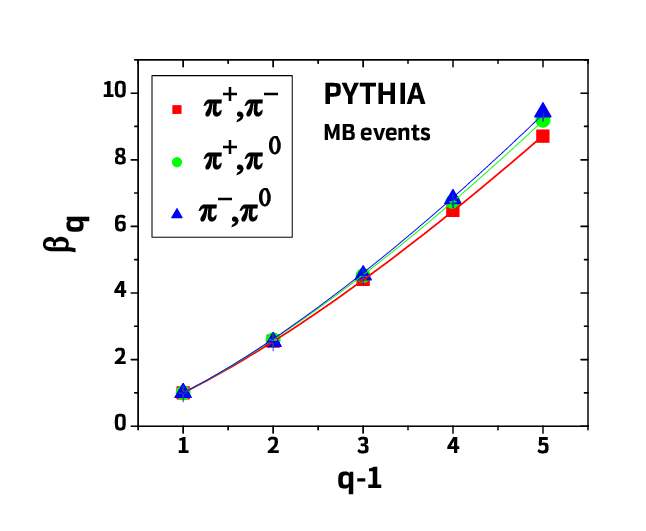}
         \caption{}
     \end{subfigure}
        \caption{\textcolor{black}{Plot of $d_q$ vs. $q$ in $\eta$-space: (a) for $\pi^+$, $\pi^-$ \& $\pi^0$ and \\ (b) for $\{\pi^+,\pi^-\}$, $\{\pi^+,\pi^0\}$ $\&$ $\{\pi^-,\pi^0\}$.}}
\end{figure}




\begin{figure}
     \centering
     \captionsetup{justification=centering}
     \begin{subfigure}[b]{0.49\textwidth}
         \centering
         \includegraphics[width=\textwidth]{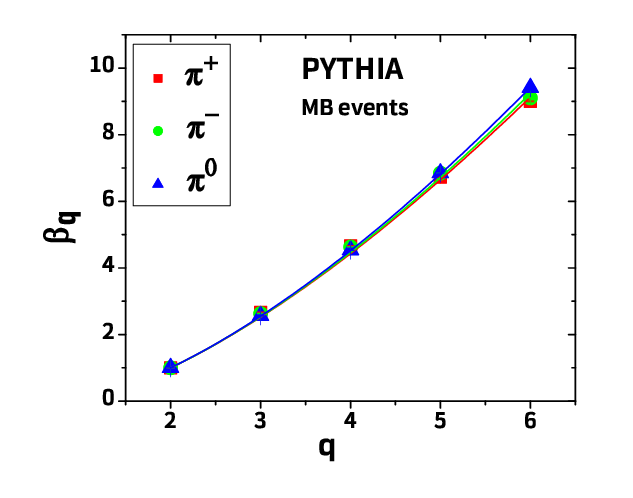}
         \caption{}
     \end{subfigure}
     \hfill
     \begin{subfigure}[b]{0.49\textwidth}
         \centering
         \includegraphics[width=\textwidth]{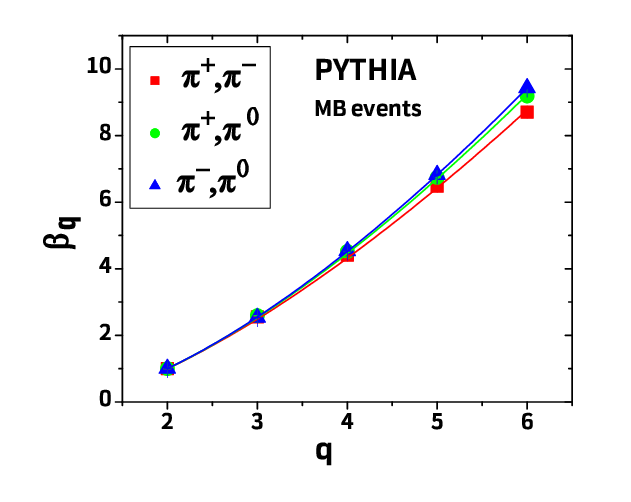}
         \caption{}
     \end{subfigure}
        \caption{\textcolor{black}{Plot of $\beta_q$ vs. $q$ in $\eta$-space: (a) for $\pi^+$, $\pi^-$ \& $\pi^0$ and \\ (b) $\{\pi^+,\pi^-\}$, $\{\pi^+,\pi^0\}$ $\&$ $\{\pi^-,\pi^0\}$.}}
\end{figure}

\setlength{\tabcolsep}{6pt} 
\renewcommand{\arraystretch}{1.25} 
\begin{table*}[htbp]
	\begin{center} \captionsetup{justification=centering}
		\caption{\textbf{The values of the various parameters associated with the SFM analysis for each types of pion generated by the PYTHIA model at MB condition within $\eta$-space.}}
		
		\resizebox{1.1\textwidth}{!}{
		
		\begin{tabular}{|c|c|c|c|c|c|c|c|}		
		
		\hline \hline
\textbf{\vtop{\hbox{\strut Elimentary}\hbox{\strut   \,\,\,Particles}}} & \textbf{Order} & \boldmath$\alpha_q$ & \boldmath$\lambda_q$ & \boldmath$d_q$ & \boldmath$r$ & \boldmath$\beta_q$ & \boldmath$\nu$ \\\hline
      \hline
		
		 & 2 & 0.018$\pm$0.005 & 0.509$\pm$0.003 & 0.018$\pm$0.005 &  & 1.000$\pm$0.010 &  \\
			& 3 & 0.048$\pm$0.009 & 0.349$\pm$0.003 & 0.024$\pm$0.005 & & 2.667$\pm$0.010 &  \\
		\boldmath$\pi^+$	& 4 & 0.084$\pm$0.012 & 0.271$\pm$0.003 & 0.028$\pm$0.004 & 0.195$\pm$0.003 & 4.667$\pm$0.009 & 1.371$\pm$0.005 \\
			& 5 & 0.121$\pm$0.017 & 0.224$\pm$0.003 & 0.030$\pm$0.004 & & 6.722$\pm$0.009 &  \\
			& 6 & 0.162$\pm$0.026 & 0.194$\pm$0.004 & 0.032$\pm$0.005 & & 9.000$\pm$0.010 &  \\\hline
      \hline		
		
		& 2 & 0.019$\pm$0.005 & 0.510$\pm$0.003 & 0.019$\pm$0.005 &  & 1.000$\pm$0.010 &  \\
			& 3 & 0.050$\pm$0.009 & 0.350$\pm$0.003 & 0.025$\pm$0.005 & & 2.632$\pm$0.010 &  \\
		\boldmath$\pi^-$	& 4 & 0.088$\pm$0.012 & 0.272$\pm$0.003 & 0.029$\pm$0.004 & 0.204$\pm$0.003 & 4.632$\pm$0.009 & 1.378$\pm$0.004 \\
			& 5 & 0.130$\pm$0.017 & 0.226$\pm$0.003 & 0.033$\pm$0.004 & & 6.842$\pm$0.009 &  \\
			& 6 & 0.173$\pm$0.024 & 0.196$\pm$0.004 & 0.035$\pm$0.005 & & 9.105$\pm$0.010 &  \\\hline
      \hline
		
		& 2 & 0.032$\pm$0.005 & 0.516$\pm$0.003 & 0.032$\pm$0.005 &  & 1.000$\pm$0.010 &  \\
			& 3 & 0.082$\pm$0.009 & 0.361$\pm$0.003 & 0.041$\pm$0.005 & & 2.563$\pm$0.010 &  \\
		\boldmath$\pi^0$	& 4 & 0.145$\pm$0.012 & 0.286$\pm$0.003 & 0.048$\pm$0.004 & 0.220$\pm$0.003 & 4.531$\pm$0.009 & 1.390$\pm$0.003 \\
			& 5 & 0.219$\pm$0.017 & 0.244$\pm$0.003 & 0.055$\pm$0.004 & & 6.844$\pm$0.009 &  \\
			& 6 & 0.301$\pm$0.025 & 0.217$\pm$0.004 & 0.060$\pm$0.005 & & 9.406$\pm$0.010 &  \\\hline
      \hline		
		
		\end{tabular}}
	\end{center}
\end{table*}

\setlength{\tabcolsep}{6pt} 
\renewcommand{\arraystretch}{1.25} 
\arrayrulecolor{black}
\begin{table*}[htbp]
	\begin{center}
	\captionsetup{justification=centering}
		\caption{\textbf{\textcolor{black}{The values of the various parameters associated with the SFM analysis for pions generated by the PYTHIA model at MB condition within $\eta$-space.}}}
		
		\resizebox{1.1\textwidth}{!}{
		
		\begin{tabular}{|c|c|c|c|c|c|c|c|}		
		
		\hline \hline
\textbf{\vtop{\hbox{\strut Elimentary}\hbox{\strut   \,\,\,Particles}}} & \textbf{Order} & \boldmath$\alpha_q$ & \boldmath$\lambda_q$ & \boldmath$d_q$ & \boldmath$r$ & \boldmath$\beta_q$ & \boldmath$\nu$ \\\hline
      \hline
		
		 & 2 & 0.027$\pm$0.005 & 0.514$\pm$0.003 & 0.027$\pm$0.005 &  & 1.000$\pm$0.010 &  \\
			& 3 & 0.069$\pm$0.008 & 0.356$\pm$0.003 & 0.035$\pm$0.004 & & 2.556$\pm$0.010 &  \\
		\boldmath$\pi^+,\pi^-$	& 4 & 0.119$\pm$0.011 & 0.280$\pm$0.003 & 0.040$\pm$0.004 & 0.182$\pm$0.003 & 4.407$\pm$0.009 & 1.346$\pm$0.001 \\
			& 5 & 0.174$\pm$0.016 & 0.238$\pm$0.003 & 0.048$\pm$0.004 & & 6.722$\pm$0.009 &  \\
			& 6 & 0.235$\pm$0.023 & 0.211$\pm$0.004 & 0.053$\pm$0.004 & & 9.000$\pm$0.010 &  \\\hline
      \hline		
		
		& 2 & 0.027$\pm$0.005 & 0.514$\pm$0.003 & 0.027$\pm$0.005 &  & 1.000$\pm$0.010 &  \\
			& 3 & 0.069$\pm$0.007 & 0.357$\pm$0.003 & 0.035$\pm$0.004 & & 2.593$\pm$0.009 &  \\
		\boldmath$\pi^+,\pi^0$	& 4 & 0.121$\pm$0.011 & 0.281$\pm$0.003 & 0.041$\pm$0.004 & 0.206$\pm$0.003 & 4.519$\pm$0.009 & 1.377$\pm$0.001 \\
			& 5 & 0.182$\pm$0.016 & 0.236$\pm$0.003 & 0.046$\pm$0.004 & & 6.741$\pm$0.009 &  \\
			& 6 & 0.248$\pm$0.022 & 0.208$\pm$0.004 & 0.050$\pm$0.005 & & 9.185$\pm$0.010 &  \\\hline
      \hline
		
		& 2 & 0.027$\pm$0.005 & 0.514$\pm$0.003 & 0.028$\pm$0.005 &  & 1.000$\pm$0.010 &  \\
			& 3 & 0.071$\pm$0.007 & 0.357$\pm$0.003 & 0.036$\pm$0.004 & & 2.536$\pm$0.009 &  \\
		\boldmath$\pi^-,\pi^0$	& 4 & 0.127$\pm$0.011 & 0.282$\pm$0.003 & 0.042$\pm$0.004 & 0.220$\pm$0.003 & 4.536$\pm$0.009 & 1.390$\pm$0.004 \\
			& 5 & 0.191$\pm$0.016 & 0.238$\pm$0.003 & 0.048$\pm$0.004 & & 6.821$\pm$0.009 &  \\
			& 6 & 0.264$\pm$0.022 & 0.211$\pm$0.004 & 0.053$\pm$0.004 & & 9.429$\pm$0.009 &  \\\hline
      \hline		
		
		\end{tabular}}
	\end{center}
\end{table*}


\setlength{\tabcolsep}{10pt} 
\renewcommand{\arraystretch}{1.25} 
\arrayrulecolor{black}
\begin{table}[h!]
	\begin{center}
	\captionsetup{justification=centering}
		\caption{\textbf{Values of the Levy Index ($\mu$)  and Specific Heat ($c$) for $\pi^+$, $\pi^-$ \& $\pi^0$ generated by the PYTHIA model at MB condition within $\eta-$space.}}
		\begin{tabular}{|c|c|c|c|c|}
		\hline \hline
		\textbf{Elementary} & \bf{Levy Index } & \textbf{Specific Heat } \\
		\textbf{Particles} & \boldmath($\mu$) & \boldmath($c$) \\\hline
		\hline		
		\boldmath$\pi^+$	& 1.281$\pm$0.023 & 0.042$\pm$0.018 \\
		\boldmath$\pi^-$ & 1.301$\pm$0.020 & 0.047$\pm$0.018\\
		\boldmath$\pi^0$ & 1.332$\pm$0.002 & 0.083$\pm$0.018\\
		\hline \hline
		
		\end{tabular}
	\end{center}
\end{table}

\setlength{\tabcolsep}{10pt} 
\renewcommand{\arraystretch}{1.25} 
\arrayrulecolor{black} 
\begin{table}[h!]
	\begin{center}
	\captionsetup{justification=centering}
		\caption{\textbf{\textcolor{black}{Values of the Levy Index ($\mu$)  and Specific Heat ($c$) for $\{\pi^+,\pi^-\}$, $\{\pi^+,\pi^0\}$ \& $\{\pi^-,\pi^0\}$ generated by the PYTHIA model at MB condition within $\eta-$space.}}}
		\begin{tabular}{|c|c|c|c|c|}
		\hline \hline
		\textbf{\textcolor{black}{Elementary}} & \bf{\textcolor{black}{Levy Index} } & \textbf{\textcolor{black}{Specific Heat} } \\
		\textbf{\textcolor{black}{Particles}} & \textcolor{black}{\boldmath($\mu$)} & \textcolor{black}{\boldmath($c$)} \\\hline
		\hline		
		\textcolor{black}{\boldmath$\pi^+,\pi^-$}	& \textcolor{black}{1.214$\pm$0.011} & \textcolor{black}{0.061$\pm$0.018} \\
		\textcolor{black}{\boldmath$\pi^+,\pi^0$} & \textcolor{black}{1.298$\pm$0.007} & \textcolor{black}{0.067$\pm$0.018}\\
		\textcolor{black}{\boldmath$\pi^-,\pi^0$} & \textcolor{black}{1.333$\pm$0.002} & \textcolor{black}{0.076$\pm$0.017}\\
		\hline \hline
		
		\end{tabular}
	\end{center}
\end{table}

Statistical errors have been included in the graphs in the form of error bars. We have performed a linear fit to the graphical points to derive the intermittency exponent $\alpha_q$ for various orders of  $q$ under the MB condition. The resulting values for PYTHIA under the MB condition are provided in Table~1 \textcolor{black}{ and Table~2}. Upon examination of the Table~1 \textcolor{black}{ and Table~2}, it becomes apparent that the  $\alpha_q$ values exhibit an increase with the order of $q$ for the charged particle multiplicities pertaining to each type of pions -- namely, $\pi^+$, $\pi^-$, and $\pi^0$ and for their mixture pairs also. This observation implies that there is no indication supporting the presence of a `non-thermal phase transition' in the case of $\pi^+$, $\pi^-$, $\pi^0$  \textcolor{black}{and also for pion pair sets of $\{\pi^{+},\pi^{-}\}$, $\{\pi^{+},\pi^{0}\}$ $\&$ $\{\pi^{-},\pi^{0}\}$. However, from Table~1 it is visible that for $\pi^+$ and $\pi^-$ the values of $\alpha_q$ are nearly equal compared to the values of $\pi^0$. The values of $\pi^0$ are larger but from Table~2 the values of $\alpha_q$ for each mixture pair set are almost same.}

 We have further calculated the values of the anomalous fractal dimension ($d_q$) in 1-dimensional $\eta$-space for each type of pions \textcolor{black}{and their mixture pairs} utilizing the intermittency exponent $\alpha_q$ according to Eqn.~3. A closer look at the variations of $d_q$, which varies linearly with $q$ for each type of pions, i.e., for $\pi^+$, $\pi^-$ and $\pi^0$, which is illustrated in Fig.~5(a) \textcolor{black}{and also for pion pair sets of $\{\pi^{+},\pi^{-}\}$, $\{\pi^{+},\pi^{0}\}$ $\&$ $\{\pi^{-},\pi^{0}\}$ which is illustrated in Fig.~5(b). Also from Fig.~5(a) it is apparent that the variation of $\pi^0$ with respect to $q$ is larger than the variations from $\pi^+$ and $\pi^-$ but for the mixture pairs the variations are almost same for every pair. } 

In this study, we have computed the values of $\beta_q$ for each type of pions, $\pi^+$, $\pi^-$, $\pi^0$, \textcolor{black}{and for pion pair sets of $\{\pi^{+},\pi^{-}\}$, $\{\pi^{+},\pi^{0}\}$ $\&$ $\{\pi^{-},\pi^{0}\}$} in 1-dimensional $\eta$-space. These values were derived for PYTHIA-generated MB events, and is presented in Table~1 \textcolor{black}{and Table~2}. Additionally, we have plotted the variation of $\beta_q$ with  $q-1$ and determined the critical exponent ($\nu$) for each type of pions separately \textcolor{black}{and for mixture pairs also}. This variation is graphically illustrated in Fig.~6(a) \textcolor{black}{and Fig.~6(b),} for PYTHIA-generated MB events. \textcolor{black}{From Fig.~6(a) and Fig.~6(b) it is apparent that for individual pions and the mixture pairs the variations $\beta_q$ with respect to $q-1$ are almost same. However, there is a little deflection can be seen at the very end of $\{\pi^+,\pi^- \}$ compared to the other two mixture sets in Fig.~6(b). }

For the PYTHIA-generated events under the MB condition, the values of $\nu$ for $\pi^+$, $\pi^-$, $\pi^0$ \textcolor{black}{and for pion pair sets of $\{\pi^{+},\pi^{-}\}$, $\{\pi^{+},\pi^{0}\}$ $\&$ $\{\pi^{-},\pi^{0}\}$ are 
incorporated in Table~1 and Table~2}. Interestingly, it should be noted that all of these critical exponent ($\nu$) values are almost close to the GL theory estimated value ($1.304$). For MB datasets produced by PYTHIA, this alignment implies the existence of the hadron-quark phase transition in 1-dimensional $\eta$-space \textcolor{black}{for both individual pions and the mixture pairs.}

We have also examined the fluctuations of $\beta_q$ concerning $q$ and computed the Levy index $(\mu)$ for each type of pions  \textcolor{black}{and mixture of pion pairs} in 1-dimensional $\eta$-space for events generated by the PYTHIA model under MB condition. This is illustrated in \textcolor{black}{Fig.~7(a) and Fig.~7(b)} . It should be noted that $\mu>1.0$ suggests the occurrence of wild singularities coming from non-Poisson like oscillations in the density distribution. 




\begin{figure}
     \centering
     \begin{subfigure}[b]{0.49\textwidth}
         \centering
         \includegraphics[width=\textwidth]{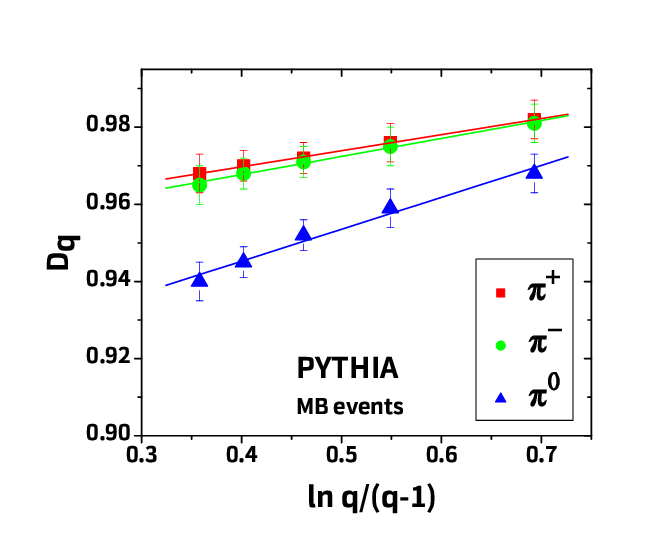}
         \caption{}
     \end{subfigure}
     \hfill
     \begin{subfigure}[b]{0.49\textwidth}
         \centering
         \includegraphics[width=\textwidth]{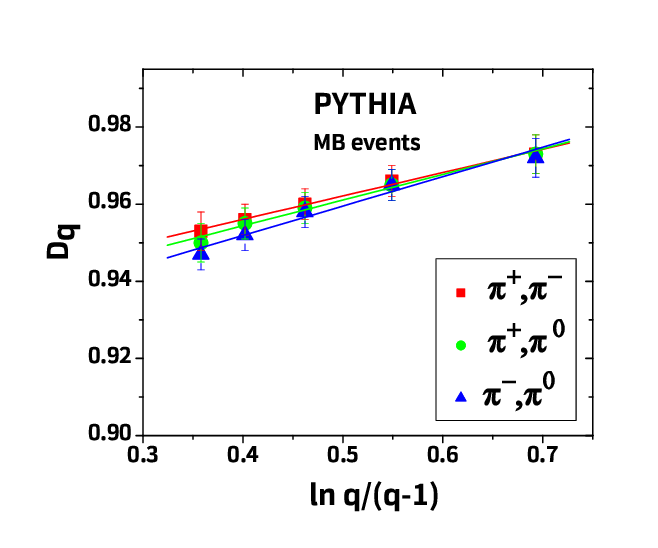}
         \caption{}
     \end{subfigure}
        \caption{\textcolor{black}{Plot of $D_q$ vs. $\ln q/(q-1)$ in $\eta$-space: (a) for $\pi^+$, $\pi^-$ \& $\pi^0$ and (b) $\{\pi^+,\pi^-\}$, $\{\pi^+,\pi^0\}$ $\&$ $\{\pi^-,\pi^0\}$.  The lines joining the data points are the best fit lines.}}
\end{figure}

\textcolor{black}{Notably, for the PYTHIA-generated events under the MB condition, the values of $\mu$  
for both individual pions and the mixture pairs, all the values of $\mu$ exceed unity, this clearly suggests that a non-thermal phase transition occurred during the cascade process. For 1-dimensional $\eta$-space, all values of $\mu$ are listed in Table~3 and Table~4.}\\

In order to understand the significance of multifractality and fractal characteristics in stochastic systems, we have used a thermodynamical perspective. The constant heat approximation is very useful for many applications in thermodynamics. \textcolor{black}{Fig.~8(a) and Fig.~8(b)} show the variation of the generalized fractal dimension $D_q$ with respect to $\left(\frac{\ln q}{q-1}\right)$. Next, we have calculated the `multifractal specific heat' $(c)$ for events generated by the PYTHIA-generated MB events in 1-dimensional $\eta$-space for \textcolor{black}{both individual pions and the mixture pairs} correspondingly.\\





\begin{figure}
     \centering
     \captionsetup{justification=centering}
     \begin{subfigure}[b]{0.49\textwidth}
         \centering
         \includegraphics[width=\textwidth]{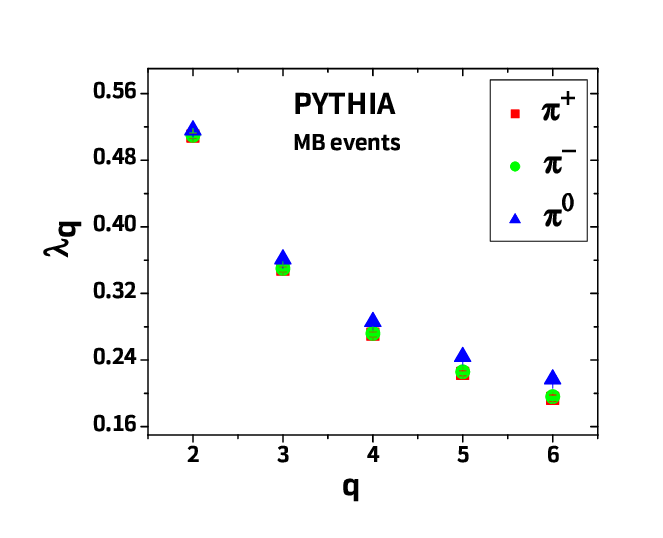}
         \caption{}
     \end{subfigure}
     \hfill
     \begin{subfigure}[b]{0.49\textwidth}
         \centering
         \includegraphics[width=\textwidth]{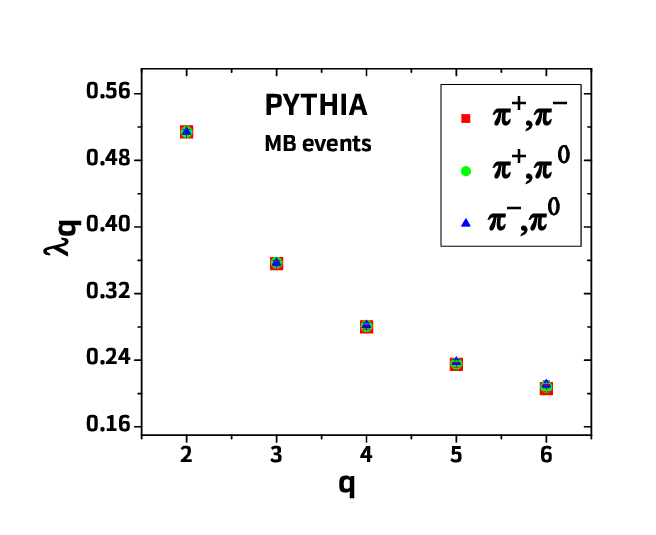}
         \caption{}
     \end{subfigure}
        \caption{\textcolor{black}{Plot of $\lambda_q$ vs. $q$ in $\eta$-space: (a) for $\pi^+$, $\pi^-$ \& $\pi^0$ and \\ (b) $\{\pi^+,\pi^-\}$, $\{\pi^+,\pi^0\}$ $\&$ $\{\pi^-,\pi^0\}$.}}
\end{figure}




\begin{figure}
     \centering
     \begin{subfigure}[b]{0.49\textwidth}
         \centering
         \includegraphics[width=\textwidth]{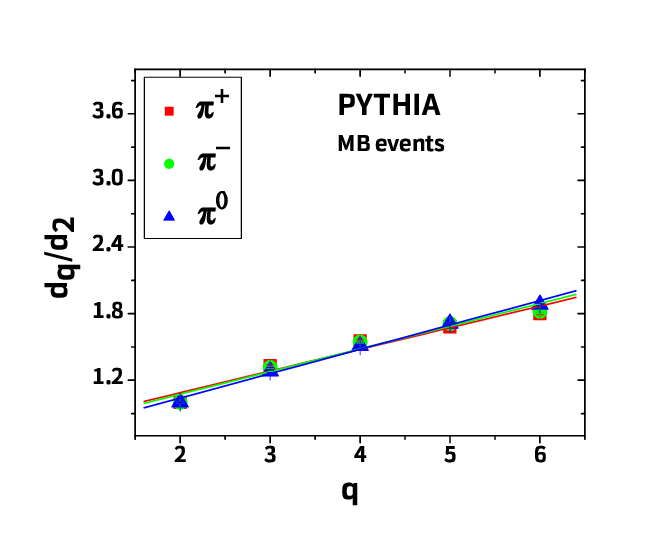}
         \caption{}
     \end{subfigure}
     \hfill
     \begin{subfigure}[b]{0.49\textwidth}
         \centering
         \includegraphics[width=\textwidth]{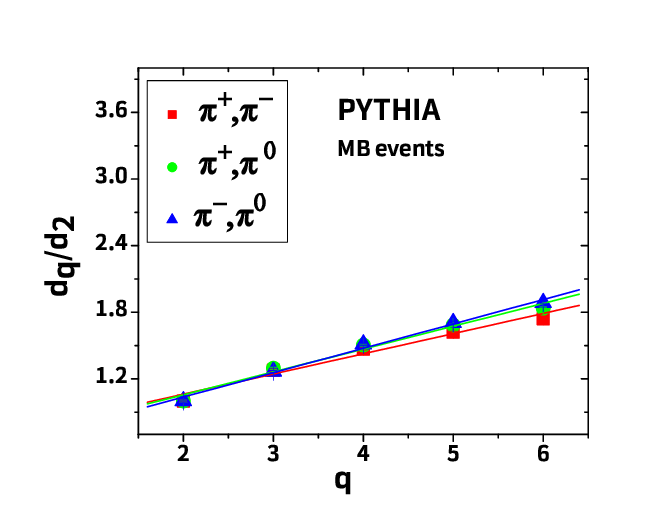}
         \caption{}
     \end{subfigure}
        \caption{\textcolor{black}{Plot of $d_q/d_2$ vs. $q$ in $\eta$-space: (a) for $\pi^+$, $\pi^-$ \& $\pi^0$ and (b) $\{\pi^+,\pi^-\}$, $\{\pi^+,\pi^0\}$ $\&$ $\{\pi^-,\pi^0\}$.  The lines joining the data points are the best fit lines.}}
\end{figure}

\textcolor{black}{In 1-dimensional $\eta$-space, for PYTHIA generated MB events, the values of multifractal specific heat ($c$) are 
interestingly found to be greater than zero for all data sets in $\eta$-space. These values show that the multiplicities of the individual pions and their mixture pairs,} generated from $pp$ collisions at an energy of $\sqrt{s}=13$ TeV, transition from multifractality to monofractality. For all data sets the values of the multifractal specific heat $(c)$ are listed in Table~3 \textcolor{black}{and Table~4} for PYTHIA generated MB events.\\

To observe the non-thermal phase transition in the charged particle multiplicities, we have calculated the values of $\lambda_q$ for various orders of $q$ for \textcolor{black}{both individual pions and their mixture pairs} in $\eta$ space. \textcolor{black}{Fig.~9(a) and Fig.~9(b)} show the change of $\lambda_q$ with respect to $q$ for PYTHIA generated MB events. \textcolor{black}{Table~1 and Table~2}  contains all of the extracted values of $\lambda_q$ for all data sets in $\eta$-space.\\

Finally, the change of the $d_q/d_2$ with the order of moment $(q)$ for \textcolor{black}{individual pions and their mixture pairs} in  $\eta$ space for PYTHIA simulated MB events are finally displayed in \textcolor{black}{Fig.~10(a) and Fig.~10(b)}. For the dataset, the degree of multifractality ($r$) can be derived by dividing the higher-order anomalous fractal dimension $d_q$ by the second-order anomalous fractal dimension $d_2$.  In $\eta$-space the resulting values are listed in \textcolor{black}{Table~1 and Table~2}. In $\eta$-space, for PYTHIA generated MB events, the degree of multifractality $(r)$ value for $\pi^0$ is greater than both $\pi^+$ and $\pi^-$, i.e., the process of generation of $\pi^0$ is more multifractal compared to $\pi^+$ and $\pi^-$. \textcolor{black}{However, the degree of multifractality $(r)$ value for $\{\pi^+,\pi^-\}$ is less than both $\{\pi^+,\pi^0\}$ and $\{\pi^-,\pi^0\}$, i.e., the process of generation of $\{\pi^+,\pi^-\}$ is less multifractal compared to $\{\pi^+,\pi^0\}$ and $\{\pi^-,\pi,^0\}$.}
\section{Conclusions}
\label{sec:cncl}

We have analyzed the distributions of the multiplicities of $\pi^+$, $\pi^-$, $\pi^0$ \textcolor{black}{along with their pairs (i.e. $\{\pi^{+},\pi^{-}\}$, $\{\pi^{+},\pi^{0}\}$ and $\{\pi^{-},\pi^{0}\}$)} in terms of fractality and phase transitions using the SFM technique for the datasets generated by the PYTHIA model under the MB condition. The key findings can be summarised as follows: -- 

\begin{itemize}

\item \textcolor{black}{As proposed earlier, it has been observed that the effects of identical particles are weakened for the mixture pion pairs (i.e. $\{\pi^{+},\pi^{-}\}$, $\{\pi^{+},\pi^{0}\}$ and $\{\pi^{-},\pi^{0}\}$) with respect to the individual distributions (i.e. $\pi^+, \pi^-, \pi^0$) which indicates the effects of BE correlations on intermittency. }

\item It has been observed that for the PYTHIA generated MB events, the values of $\nu$ are almost identical with the GL theory predicted value, which suggests the \textcolor{black}{apparent} existance of the hadron-quark phase transition in each types of pions \textcolor{black}{ and also for their mixture pairs}.
\item It has also been observed that for PYTHIA generated events at MB condition, the values of the Levy index ($\mu$) for $\pi^+$, $\pi^-$, $\pi^0$  and \textcolor{black}{the mixture pairs} are greater than unity in the case of 1-dimensional $\eta$-space which indicates the signature of non-thermal phase transition and correspond to ``wild'' singularities. 
\item In 1-dimensional $\eta$-space, the values of  $c$ for $\pi^+$, $\pi^-$, $\pi^0$  \textcolor{black}{and the mixture data sets} are greater than zero for PYTHIA generated MB events, indicate the transition from multifractality to monofractality in $pp$ collisions at $\sqrt{s}=13$ TeV.

\item For PYTHIA generated MB events, the value of the
degree of multifractality ($r$) for $\pi^0$ is greater than the both $\pi^+$ and $\pi^-$, i.e., the process of generation of $\pi^0$ is more multifractal compare to $\pi^+$ and $\pi^-$. \textcolor{black}{However, the value of $r$ for $\{\pi^+,\pi^-\}$ is less than the both $\{\pi^+,\pi^0\}$ and $\{\pi^-,\pi^0\}$, i.e., the process of generation of $\{\pi^+,\pi^-\}$ is less multifractal compare to $\{\pi^+,\pi^0\}$ and $\{\pi^-,\pi^0\}$}.

\end{itemize}

It's crucial to acknowledge that the observable demonstrated in this study to analyze fractality and phase transition may yield false signals of the quark-hadron phase transition. This susceptibility arises from random fluctuations in particle production. Nonetheless, the observable and methodology outlined can be subjected to further testing using additional heavy-ion collisions and a model that incorporates the quark-hadron phase transition. Such investigations are intriguing because they can help us better \textcolor{black}{understand} the pions emission phenomenon by providing insight into the pions emission process. The degree of \textcolor{black}{multifractality} shows its dependencies on the special distribution of each class of pions, i.e., $\pi^+$, $\pi^-$ and $\pi^0$ during the multiparticle production process. The contribution to the PYTHIA minimum bias results is derived from the two particle correlations and single particle density fluctuations. Such type of analysis will provide some light on the preferential emission \textcolor{black}{mechanism} of  specific classes of pions (i.e., $\pi^+$, $\pi^-$ and $\pi^0$) which are highly beneficial to understand the preferential emission phenomena of pions.

\section{Acknowledgements}
\textcolor{black}{The authors would like to express their gratitude to the anonymous reviewer for providing valuable feedback that has enhanced the accuracy and readability of the paper.} T. Biswas acknowledges the financial support as an Inspire fellow (No. DST/INSPIRE Fellowship/2022/IF220173) from the Department of Science and Technology, Government of India. 

\section{Conflict of Interest}

All the authors announce that they have no conflict of interest.


\end{document}